\documentclass[12pt]{article}
\usepackage{epsfig}

\newcommand{\bea}{\begin{eqnarray}}
\newcommand{\eea}{\end{eqnarray}}
\newcommand{\be}{\begin{equation}}
\newcommand{\ee}{\end{equation}}

\newcommand{\Li}{\mathop{\mathrm{Li}}\nolimits}

\title{$F_2^c$ 
at low $x$
\author{A. Yu.~Illarionov$^a$ and  A. V.~Kotikov$^b$ \\
$^a$ Dipartimento di Fisica, Universita' di Trento, Italy \\
$^b$ BLThPh,
Joint Institute for Nuclear Research, Russia} }

\begin{document}
\maketitle
\abstract{
We study the heavy-quark contributions to the proton structure function
$F_2(x,Q^2)$
at next-to-leading order using
compact formulas 
at small values of Bjorken's
$x$ variable.
The formulas
provide a good agreement with the modern HERA data for
$F_2^c(x,Q^2)$.
}\\

{\it Keywords:} Deep inelastic scattering; nucleon structure functions;
QCD coupling constant.

$PACS:~~12.38~Aw,\,Bx,\,Qk$\\

\vskip 1cm

\section{Introduction}
\label{sec:intro}

In the last year
the H1 \cite{Collaboration:2009jy,:2009ut} and ZEUS
\cite{Abramowicz:2010zq,Chekanov:2009kj}
Collaborations at HERA are presented a new data
on the charm structure function (SF) $F_2^c$
\footnote{The papers
\cite{Collaboration:2009jy,:2009ut,Abramowicz:2010zq,Chekanov:2009kj}
contain also the references on the previous data on
deep-inelastic (DIS)
structure function (SF) $F_2^c$ 
at small $x$ values.}.
Moreover, recently  H1 and ZEUS have been demonstrated
the preliminary combine data of $F_2^c(x,Q^2)$ 
\cite{Lipka:2009zza}.

In the framework of Dokshitzer-Gribov-Lipatov-Altarelli-Parisi (DGLAP)
dynamics \cite{Gribov:1972ri} there are two basic methods to study
heavy-flavor physics.
One of them \cite{Kniehl:1996we} is based on the massless evolution of parton
distribution functions (PDFs) and the other one is based 
on the photon-gluon fusion (PGF) process
\cite{Frixione:1994dv}.
There are also some interpolating schemes (see
Ref.~\cite{Olness:1987ep} and references cited therein).

In this short paper,
we apply compact low-$x$
approximations for the
SF $F_2^c(x,Q^2)$
obtained \cite{Illarionov:2008be}
in the PGF framework 
at the first two orders
of perturbation theory to these new HERA experimental data
\cite{Collaboration:2009jy,:2009ut,Abramowicz:2010zq,Chekanov:2009kj,Lipka:2009zza}.
We show a good agreement between experimental data and the approach which was
found
without any additional  free parameters. All
PDF parameters
have been fitted
earlier \cite{Illarionov:2004nw} from  $F_2(x,Q^2)$ HERA
experimental data only.

\section{Approach
}
\label{sec:approach}

We now present
our basic
formula for $F_2^c(x,Q^2)$ appropriate for small values
of $x$,
where only the gluon and quark-singlet contributions
matter, while the non-singlet contributions are negligibly small
\footnote{%
Here and in the following, we suppress the variables
$m_c$ in the
argument lists of the structure and coefficient functions for the ease of
notation.}
\begin{equation}
F_2^c(x,Q^2)=\sum_{a=g,l,\overline{l}}\sum_{r=+,-}
C_{2,a}^r(x,Q^2,\mu^2)\otimes xf_a^r(x,\mu^2) \approx \sum_{r=+,-}
C_{2,g}^r(x,Q^2,\mu^2)\otimes xf_g^r(x,\mu^2)
\label{eq:pm}
\end{equation}
where $l$ generically denotes the light-quark flavors,
$r=\pm$ labels the usual $+$ and $-$ linear combinations of the gluon
and quark-singlet contributions, $C_{2,a}^r(x)$ are the DIS coefficient
functions, which can be calculated perturbatively in the parton model of QCD,
$\mu$ is the factorization scale and same time the renormalization one
appearing in the strong-coupling constant
$\alpha_s(\mu)$,
and the symbol $\otimes$ denotes
convolution according to
the usual prescription, $f(x)\otimes g(x)=\int_x^1(dy/y)f(y)g(x/y)$.
Massive kinematics requires that $C_{2,a}^r=0$ for $x>b=1/(1+4c)$, where
$c=m_c^2/Q^2$.
We take $m_c$ to be the solution of $\overline{m}_c(m_c)=m_c$, where
$\overline{m}_c(\mu)$ is defined in the modified minimal-subtraction
($\overline{\mathrm{MS}}$) scheme.
The
simplification in the r.h.s. of (\ref{eq:pm})
is obtained by neglecting the contributions due to
incoming light quarks and antiquarks, which is justified
because they vanish at LO and are numerically suppressed at NLO for small
values of $x$.

Exploiting the low-$x$ asymptotic behavior of $f_a^r(x,Q^2)$
\cite{Lopez:1979bb,Kotikov:1993xe},
\begin{equation}
f_a^r(x,\mu^2)\stackrel{x\to0}{\to}\frac{1}{x^{1+\delta_r}}
\tilde{f}_a^r(x,\mu^2)\qquad
(\mbox{hereafter}~~ a=q,g),
\label{eq:pm0}
\end{equation}
where the rise of $\tilde{f}_a^r(x,\mu^2)$ as $x\to0$ is less than any power of
$x$, Eq.~(\ref{eq:pm}) can be rewritten as
\begin{equation}
F_2^c(x,Q^2)\approx\sum_{r=+,-}
M_{2,g}^r(1+\delta_r,Q^2,\mu^2) \, xf_g^r(x,\mu^2),
\label{eq:pm1}
\end{equation}
where
\begin{equation}
M_{2,g}^r(n)=\int_0^{b_i}dx\,x^{n-2}C_{2,g}^r(x)
\label{eq:mel}
\end{equation}
is the Mellin transform, which is to be analytically continued from integer
values $n$ to real values $1+\delta_r$.
Strictly speaking, the equation (\ref{eq:pm1}) is correct for non-singular
Mellin moments $M_{2,g}^r(n)$ at $n \to 1$. The generalization of (\ref{eq:pm1})
for the moments containing singularity will be done in subsection 4.2.

\section{Gluon density
}
\label{sec:gluon}

As demonstrated in Refs.~\cite{Kotikov:1998qt,Illarionov:2004nw}, HERA data
for $F_2(x,Q^2)$ support the modified Bessel-like behavior of PDFs at small $x$ values
predicted in the framework of the so-called generalized double-asymptotic
scaling regime.
\footnote{It is a generalization of earlier studies \cite{De Rujula:1974rf}.}
In this approach,
DGLAP dynamics \cite{Gribov:1972ri} starting at some
initial value $\mu^2_0$ with flat $x$ distributions:
\begin{equation}
    x f_a (x,\mu^2_0) = A_a ,
\label{1}
\end{equation}
where
$A_a$ are unknown constants to be determined from the data.

In the framework of the generalized double-asymptotic
scaling regime,
the main ingredients of the results for gluon density
 include the following.\footnote{
The
results for quark densities
may be found in Refs.~\cite{Kotikov:1998qt,Illarionov:2004nw}.}
It is presented in terms of two
components (``$+$" and ``$-$")
\begin{equation}
    f_g(x,\mu^2)=f_g^{+}(x,\mu^2) + f_g^{-}(x,\mu^2),
\label{intro:1}
\end{equation}
which are obtained from the analytic $\mu^2$-dependent expressions of the
corresponding (``$+$" and ``$-$") PDF moments.
Here,  $e_i$ is the fractional electric charge of heavy quark $i$,
$e=(\sum_{i=1}^f e_i^2)/f$ is the average charge square and $f$ is the
number of active quark flavors.

The small-$x$ asymptotic results for the PDF $f^{\pm}_g$ are at the
next-to-leading order (NLO) \cite{Kotikov:2007ua}
\begin{eqnarray}
    xf^{+}_g(x,\mu^2) &=& A_g^{+}
        I_0(\sigma)  e^{-\overline d_{+}
s-\overline D_{+}
p}
+ O(\rho),
    \nonumber \\
    xf^{-}_g(x,\mu^2) &=& A_g^{-} e^{-d_{-}
s-D_{-}
p } + O(x) ,
    \label{8.02}
\end{eqnarray}
where $I_{\nu}$ are the modified Bessel functions,
\be
 D_{\pm}=d_{\pm\pm}-\frac{\beta_1}{\beta_0} d_{\pm}
   \label{8.02aa}
\ee
and similar for $\hat D_{+}$ and $\overline D_{+}$,
\be
A_g^{+} ~=~ \left(1-\frac{80f}{81}a_s(Q)\right)A_g + \frac{4}{9}
\left(1+\Bigl(3+\frac{f}{27}\Bigr)a_s(Q_0) -\frac{80f}{81}a_s(Q)\right)A_q,
\qquad A_g^{-} ~=~ A_g - A_g^{+}
\label{8.02a}
\ee
and
\begin{eqnarray}
a_s(\mu)&=&\frac{\alpha_s(\mu)}{4\pi},\qquad
\hat d_{+} = - \frac{12}{\beta_0},\qquad
\overline d_{+}
= 1 + \frac{20f}{27\beta_0},\qquad
d_{-}
= \frac{16f}{27\beta_0},\nonumber \\
\hat d_{++}&=& \frac{412f}{27\beta_0},\qquad
\overline d_{++}
= \frac{8}{\beta_0} \left(
36\zeta_3+33\zeta_2-\frac{1643}{12} + \frac{2f}{9}
\left[\frac{68}{9}-4\zeta_2-\frac{13f}{243}\right]\right), \nonumber \\
d_{--}
&=& \frac{16}{9\beta_0} \left(
2\zeta_3-3\zeta_2+\frac{13}{4} + f
\left[4\zeta_2-\frac{23}{18}+\frac{13f}{243}\right]\right),
\label{8.02b}
\end{eqnarray}
with $\zeta_3$ and $\zeta_2$ are Euler functions.
$\beta_0$ and $\beta_1$ are first two coefficients of QCD $\beta$-function.

The variables $s$, $p$, $\sigma$, and $\rho$ are
given by
\begin{equation}
    s = \ln\frac{a_s(\mu_0)}{a_s(\mu)},~
p= a_s(\mu_0)-a_s(\mu),~
\sigma = 2\sqrt{\left(\hat{d}_{+}s +\hat{D}_{+}p\right) \ln x},~
    \rho = \frac{\sigma}{2\ln(1/x)}.
\label{slo}
\end{equation}

The transformation to the LO is simple: in Eqs.  (\ref{8.02}), (\ref{8.02a})
and (\ref{slo})
we should put $p=a_s(Q)=a_s(Q_0)=0$ and replace the variable $s$ by its
LO approximation:
$s^{\rm LO}=\ln(a^{\rm LO}_s(\mu_0)/a^{\rm LO}_s(\mu))$.

\section{Wilson coefficients}
\label{sec:Wilson}

One has $M_{2,g}^+(1)=M_{2,g}^-(1)$ if
$M_{2,g}^r(n)$ are devoid of singularities in the limit $\delta_r\to0$, as
we assume for the time being.
Such singularities actually occur at NLO, leading to modifications to be
discussed below.
Defining $M_{2,g}(1)=M_{2,g}^\pm(1)$ and using
$f_g(x,Q^2)=\sum_{r=\pm}f_g^r(x,Q^2)$, Eq.~(\ref{eq:pm1}) may be simplified to
become
\begin{equation}
F_2^c(x,Q^2)\approx M_{2,g}(1,Q^2,\mu^2)xf_g(x,\mu^2).
\label{eq:pm2n}
\end{equation}

Through NLO, $M_{2,g}(1,Q^2,\mu^2)$ exhibits the structure
\be
M_{2,g}(1,Q^2,\mu^2) = e_i^2a_s(\mu)\left\{M_{2,g}^{(0)}(1,c)
+a_s(\mu)\left[M_{2,g}^{(1)}(1,c)+M_{2,g}^{(2)}(1,c) \ln\frac{\mu^2}{m_c^2}
\right]\right\}+{\mathcal O}(a_s^3).
\label{eq:exp}
\ee


\subsection{LO results}
\label{sec:lo}

The LO coefficient function of PGF can be obtained from the QED case
\cite{Baier:1966bf} by adjusting coupling constants and color factors, and
they read \cite{Witten:1975bh,Kotikov:2001ct}
\be
C_{2,g}^{(0)}(x,c) ~=~ -2x\{[1-4x(2-c)(1-x)]\beta
-[1-2x(1-2c)
+2x^2(1-6c-4c^2)]L(\beta)\},
\label{eq:lo}
\ee
where
\begin{equation}
\beta(x)=\sqrt{1-\frac{4cx}{1-x}},\qquad
L(\beta)=\ln\frac{1+\beta}{1-\beta}.
\label{eq:lo1}
\end{equation}

Performing
the Mellin transformation \cite{Illarionov:2008be} in Eq.~(\ref{eq:mel}),
we find
\be
M_{2,g}^{(0)}(1,c) ~=~ \frac{2}{3}[1+2(1-c)J(c)]
\label{eq:lo2}
\ee
with
\be
J(c) = - \sqrt{b}\ln t,\qquad t=\frac{1-\sqrt{b}}{1+\sqrt{b}},\qquad
b=\frac{1}{1+4c}.
\label{eq:lo2a}
\ee

\subsection{NLO results}
\label{sec:nlo}

The NLO coefficient functions of PGF are rather lengthy and not published in
print; they are only available as computer codes \cite{Laenen:1992zk}.
For the purpose of this letter, it is sufficient to work in the high-energy
regime, defined by $x\ll1$, where they assume the compact form
\cite{Catani:1992zc}
\begin{equation}
C_{2,g}^{(j)}(x,c)=\beta R_{2,g}^{(j)}(1,c),
\label{eq:nlo}
\end{equation}
with
\be
R_{2,g}^{(1)}(1,c)~=~\frac{8}{9}C_A[5+(13-10c)J(c)+6(1-c)I(c)],\qquad
R_{2,g}^{(2)}(1,c)~=~-4 C_A M_{2,g}^{(0)}(1,c),
\label{eq:nloA}
\ee
where $C_A=N$ for the color gauge group SU(N), $J(c)$ is defined by
Eq.~(\ref{eq:lo2a}), and
\begin{equation}
I(c)=-\sqrt{b}\left[\zeta(2)+\frac{1}{2}\ln^2t-\ln(bc)\ln t+2\Li_2(-t)\right],
\label{eq:nlo1}
\end{equation}
where $t$ is given in (\ref{eq:lo2a}) and
$\Li_2(x)=-\int_0^1(dy/y)\ln(1-xy)$ is the dilogarithmic function.

As already mentioned above (see the end of Section 2), the Mellin transforms of
$C_{k,g}^{(j)}(x,c)$ exhibit singularities in the limit $\delta_r\to0$, which
lead to modifications in Eqs.~(\ref{eq:pm1}) and (\ref{eq:pm2n}).
As was shown in Refs.~\cite{Kotikov:1993xe,Kotikov:1998qt,Illarionov:2004nw},
the terms involving $1/\delta_r$ correspond to singularities of 
the Mellin moments $M_{2,g}^r(n)$ at $n \to 1$ and
depend on the exact form of the subasymptotic
low-$x$ behavior encoded in $\tilde{f}_g^r(x,\mu^2)$.
The modification is simple:
\begin{equation}
\frac{1}{\delta_r} \to \frac{1}{\tilde \delta_r},\qquad
\frac{1}{\tilde \delta_r}=\frac{1}{\tilde{f}_g^r(\hat{x},\mu^2)}
\int^1_{\hat{x}}\frac{dy}{y}\tilde{f}_g^r(y,\mu^2),
\label{eq:nlo2}
\end{equation}
where $\hat{x}=x/b$.
In the generalized double-asymptotic scaling regime, the $+$ and $-$
components of the gluon PDF exhibit the
low-$x$ behavior
(\ref{8.02}).
We thus have \cite{Illarionov:2008be,Illarionov:2004nw}
\begin{equation}
\frac{1}{\tilde \delta_+}
\approx \frac{1}{\rho(\hat{x})}\,
\frac{I_1(\sigma(\hat{x}))}{I_0(\sigma(\hat{x}))},
\qquad
\frac{1}{\tilde \delta_-}
\approx \ln\frac{1}{\hat{x}},
\label{eq:nlo3}
\end{equation}
where $\sigma$ and $\rho$ are given in (\ref{slo}).

Because the ratio $f_g^-(x,Q^2)/f_g^+(x,Q^2)$ is rather small at the $Q^2$
values considered, Eq.~(\ref{eq:pm2n}) is modified to become
\begin{equation}
F_2^c(x,Q^2)\approx\tilde{M}_{2,g}(1,\mu^2,c)xf_g(x,\mu^2),
\end{equation}
where $\tilde{M}_{2,g}(1,\mu^2)$ is obtained from $M_{2,g}(n,\mu^2)$ by taking
the limit $n\to 1$ and replacing $1/(n-1)\to1/\tilde \delta_+$.
Consequently, one needs to substitute
\begin{equation}
M_{2,g}^{(j)}(1,c)\to\tilde{M}_{2,g}^{(j)}(1,c)\quad(j=1,2)
\end{equation}
in the NLO part of Eq.~(\ref{eq:exp}).
Using the identity
\begin{equation}
\frac{1}{I_0(\sigma(\hat{x}))}
\int^1_{\hat{x}}\frac{dy}{y}\beta\left(\frac{x}{y}\right) I_0(\sigma(y))
\approx \frac{1}{\tilde \delta_+}-\ln (bc)-\frac{J(c)}{b},
\end{equation}
we find the Mellin transform~(\ref{eq:mel}) of Eq.~(\ref{eq:nlo}) to be
\footnote{Note, that $\delta_+$ determines the behavior of the slope of
gluon density (see (\ref{eq:pm0}))
and also mostly the  slope of SF $F_2$. The form (\ref{eq:nlo3})
of $\tilde \delta_+$ is in good agreement \cite{Kotikov:2002fd}
with the corresponding HERA experimental
data.}
\begin{equation}
\tilde{M}_{2,g}^{(j)}(1,c)\approx
\left[\frac{1}{\delta_+}-\ln(bc)-\frac{J(c)}{b}\right]
R_{2,g}^{(j)}(1,c)\quad(j=1,2),
\label{eq:nloA}
\end{equation}
with $R_{2,g}^{(j)}(1,a)\quad(j=1,2)$ are given in (\ref{eq:nloA}).
The rise of the NLO terms as $x\to0$ is in agreement with earlier
investigations \cite{Nason:1987xz}.

\section{Results}
\label{sec:results}

We are now in a position to explore the phenomenological implications of our
results.
As for our input parameters, we choose
$m_c=1.25$~GeV 
in agreement with Particle Data Group
\cite{Amsler:2008zzb}.
While the LO result Eq.~(\ref{eq:lo2}) is independent of the
unphysical mass scale $\mu$, the NLO formula~(\ref{eq:exp}) does depend on
it, due to an incomplete compensation of the $\mu$ dependence of $a(\mu)$ by
the terms proportional to $\ln(\mu^2/Q^2)$, the residual $\mu$ dependence
being formally beyond NLO.
In order to fix
the theoretical uncertainty resulting from this, we put
$\mu^2=Q^2+4m_c^2$, which is the standard scale in heavy quark production.

The PDF parameters $\mu_0^2$, $A_q$ and $A_g$ shown in (\ref{1}),
(\ref{8.02}) and (\ref{8.02a}) have been fixed in the
fits of $F_2$ experimental data. Their values depend
on conditions chosen in the fits: the order of perturbation theory and the
number $f$ of active quarks.

Below $b$-quark threshold,  the scheme with $f=4$ has been used
\cite{Illarionov:2004nw} in the
fits of $F_2$ data.
Note, that the $F_2$ structure
function contains $F_2^c$ as a part. In the fits, the NLO gluon density
and the LO and NLO quark ones contribute to $F_2^c$, as the part of $F_2$.
Then, now in PGF scattering the LO coefficient function (\ref{eq:lo})
corresponds in $m \to 0$ limit to the standard NLO Wilson coefficient
(together with the product of the LO anomalous dimension $\gamma_{qg}$ and
$\ln (m^2_c/Q^2)$). It is a general situation, i.e.
the coefficient function of
PGF scattering at some order of perturbation theory corresponds to the
standard DIS Wilson coefficient
with the one step  higher order.
The reason is following:
the standard DIS
analysis starts with handbag
diagram of photon-quark scattering
and photon-gluon interaction begins
at one-loop level.

To analyze $F_2^c$ at the LO of PGF process one
should take $xf_a(x,Q^2)$ from the fit of $F_2$ at NLO with $f=4$.
In practice, we use the following parameters,
$Q_0^2=0.523$~GeV$^2$,
$A_g=0.060$ and $A_q=0.844$.

Correspondingly, to analyze
PGF process at NLO one 
needs to know the gluon density extracted
from 
$F_2$ data at NNLO,
which is not yet known
\footnote{The difficulty to extend the analysis
\cite{Kotikov:1998qt,Illarionov:2004nw} to NNLO 
is related to the
appearance of the pole $\sim 1/(n-1)^2$ in the three-loop corrections to the
anomalous dimension $\gamma_{gg}$ (see \cite{Fadin:1998py}). The pole
$\sim 1/(n-1)^2$ violates the Bessel-like solution (\ref{8.02}) of DGLAP
equation for
PDFs at low $x$ values with the flat initial condition (\ref{1}).}
in
generalized double-asymptotic scaling regime. However, as we can see from the
modern global fits \cite{Dittmar:2005ed}, the difference between NLO and
NNLO gluon
densities is not so large. So, we can safely apply the NLO form (\ref{8.02}) of
$xf_g(x,Q^2)$ to our  NLO PGF analysis.


\begin{figure}[t]
\begin{center}
\epsfig{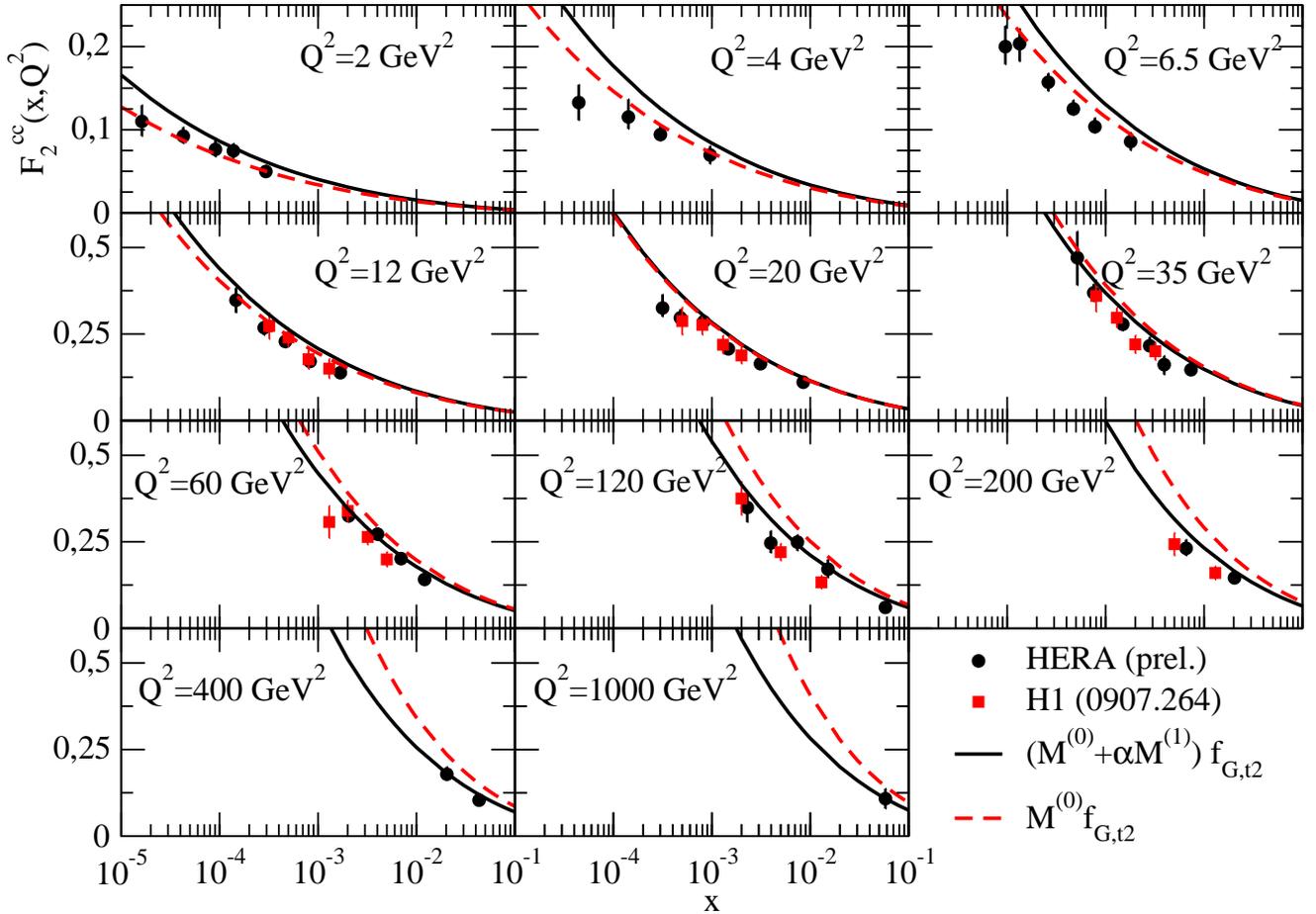}
\end{center}
\vskip 1.0cm
\caption{(colored online) The charm structure function $F_2^c(x,Q^2)$ 
evaluated as function of $x$ with the LO
matrix elements from Eq.~(\ref{eq:lo2}) (dashed lines) and with the NLO ones
from Eq.~(\ref{eq:nloA}) and
with the factorization/renormalization scale $\mu^2=Q^2+4m_c^2$ (solid lines).
The black points and red squares correspond to the combine 
preliminary H1-ZEUS data \cite{Lipka:2009zza} and H1 data
 \cite{Collaboration:2009jy,:2009ut}, respectively.
}
\label{fig:r}
\end{figure}

The results for $F_2^c$ 
are presented in Fig.1.
We can see a good agreement between our compact
formulas 
(\ref{eq:pm2n}), (\ref{eq:lo2}) and (\ref{eq:nloA})
and the modern experimental data
\cite{Collaboration:2009jy,:2009ut,Abramowicz:2010zq,Chekanov:2009kj,Lipka:2009zza}
for $F_2^c(x,Q^2)$  structure function.
To keep place on Fig.1, we show only the H1 \cite{Collaboration:2009jy,:2009ut}
 and the combine preliminary H1-ZEUS \cite{Lipka:2009zza} data.

The good agreement between generalized double-asymptotic
scaling approach 
and $F_2$,
$F_2^c$ 
data demonstrates an equal importance
of the both parton densities (gluon and sea quark) at low $x$.
It is due to the fact that $F_2$ 
relates 
mostly to the sea quark distribution, while the $F_2^c$ relates mostly to the
 gluon one. 
Dropping sea quarks in analyze leads to the different gluon densities 
extracted from $F_2$ or from $F_2^c$ (see, for example, \cite{Jung:2007qh}).

\section{Conclusions}
\label{sec:conclusions}

We presented a compact formulas for
the heavy-flavor contributions to the proton structure functions $F_2$
valid through NLO at small values of Bjorken's $x$ variable.
Our results agree with modern experimental data
\cite{Collaboration:2009jy,:2009ut,Abramowicz:2010zq,Chekanov:2009kj,Lipka:2009zza}
well within errors
without a free additional parameters.
In the $Q^2$ range probed by the HERA data, our NLO predictions agree quite 
well with the LO ones.
Since we worked in the fixed-flavor-number scheme, our results are bound to
break down for $Q^2\gg4m_c^2$, which manifests itself by appreciable QCD
correction factors and scale dependence.
As is well known, this problem is conveniently solved by adopting the
variable-flavor-number scheme, which we leave for future work.

\section*{Acknowledgements}

A.V.K. was supported in part by the Alexander von Humboldt Foundation.
The work was supported in part by RFBR grant No.10-02-01259-a.

Calculations were partially performed on the HPC facility of SISSA/Democritos in Trieste
and partially on the HPC facility ``WIGLAF'' of the Department of Physics,
University of Trento.

\end{document}